# Agent based Model for providing optimized, synchronized and failure free execution of workflow process.

Sohail Safdar[1], Dr. Jamil Ahmad[2], Shaftab Ahmed[3], M. Tayyab Asghar[4], Saqib Saeed[5]

[1]Department of Computer Science & Engineering, Bahria University, Islamabad, Pakistan
[2]Department of Computer Science, Iqra University, Islamabad, Pakistan
[3]Department of Computer Science & Engineering, Bahria University, Islamabad, Pakistan
[4]Prosol, Pi Sigma Groups, Islamabad, Pakistan
[5]Department of Computer Science & Engineering, Bahria University, Islamabad, Pakistan
Corresponding email: sagi_636@yahoo.com

**Abstract**:
The main objective of this paper is to provide an optimized solution and algorithm for the execution of a workflow process by ensuring the data consistency, correctness, completeness among various tasks involved. The solution proposed provides a synchronized and failure free flow of execution among various tasks involved in a workflow process. A synchronizing agent is bound at a very low level, i.e. with the workflow activity or task to get the desired goals to be done and an algorithm is provided to show the execution of workflow process completely.

## I. Introduction:

*Workflow Management Systems* are those systems that are used to automate, manage, monitor and control the execution of the workflow processes [1,8]. *Workflow process* is a business process for an enterprise. Enterprise has different departments and each department carries out some set of required processes to ensure the successful completion of the whole work and achieve the required goals. A process is a composition of a set of related activities known as *workflow activities or tasks*. The processes as well as the composed tasks are used to share information among each other for their individual execution. These processes and tasks may reside at same location or may be at some remote location as in the case of distributed environment.

The workflows in the distributed environment must provide correctness and consistency of data [3] that is shared among different tasks. For achieving the required business goal, the execution of the distributed workflows should be made failure free. Workflow task is the focal point in the current research, as the successful execution of the task will ensure the success of the whole workflow process.

The workflow management systems are very compelling area of study in the enterprise perspective. *Workflow processes* have to be defined for business activities and their successful execution is done to accomplish the overall goal for that enterprise. Following diagram shows the basic workflow process.

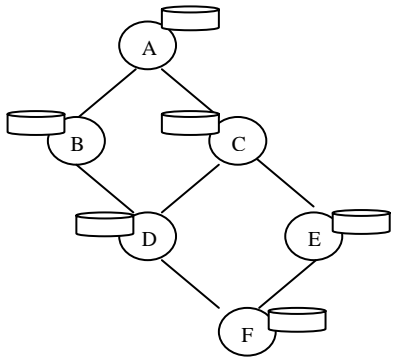

*Figure 1a: Basic workflow process*

In the above mentioned figure A, B, C, D, E and F are the tasks or workflow activities that need to be executed to complete the workflow process. In case of distributed environment each task may be located at different location, hence the local storage is provided at each task level.

The objectives of the research are the identification of key issues regarding process synchronization, data correctness, completeness and consistency among different tasks associated with the distributed workflow process. Analyzing the existing solutions to extract the idea for the possible optimized solution that will provide failure preventive execution of workflow process by ensuring data consistency, data correctness and task synchronization. To develop an algorithm for the proposed solution mentioned above.

**Organization of the paper**: The rest of the paper is organized as follows.
Section II discusses related works and useful concepts in workflow management systems. In section III the identified issues regarding consistency and correctness of the data or information are discussed. The proposed solution and its application to the existing workflow tasks are discussed in section IV. In section V the algorithm for the proposed solution is discussed. The conclusion will be drawn from the research and limitations along with the future work will be explained afterwards.

## II. Related Work and Useful Concepts:

Various researchers are working on workflow management systems. Few terminologies need to be discussed before focusing to the actual context of the research.

**Workflow Management Systems:** The workflow management systems are used to automate, manage, monitor and control the execution of the workflows in different business domains.

Different approaches are followed in workflow management systems i.e. Active Database Management Systems that act upon changes of persistent but its weakness lies in maintaining the consistent nature when the event occurs, Component based systems is based on small well defined components written in ActiveX, Agent Technology is the third approach and is the most intelligent approach, in which different agents collaborate together for the successful execution of the workflows in distributed environment [1, 8].

**Components of Distributed Workflows:**

*Workflow Process Specification*: A predefined set of tasks to be carried out in a specific workflow process.
*Workflow Server*: The server that responds to events by initiating the respective workflows according to the process specification hence manages, monitors and control overall workflow.
*Workflow Process*: A process that defines a business activity.
***Task*:** Also known as workflow activity that is a component of a workflow process. The task may consist of single or multiple instructions.
***Actor*:** Also known as workflow participant, it might be a human user or an application or another task that performs a workflow activity.
***Worklist*:** List that contains a set of activities to be performed and is handed over to the workflow handler.
***Workflow Agent*:** It is a procedure that takes the worklists from the workflow server and sends them to workflow handler.
***Workflow Handler*:** It is software that interacts with the end-user through Worklist browser and plug-ins to execute the workflow activity.

***Workflow Engine*:** It is a service that provides the execution environment for the workflow process.
***Resource Manager*:** The service that provides the required resources on demand. It works along with workflow handler.
***Worklist browser and plug-in*:** This works as an interface between user and workflow tasks. [1, 2]

The predefined workflow processes in distributed environment can be initiated through workflow server using agents that will give the handle of worklist, to the workflow handlers which along with resource manager and workflow engine results in the execution of that workflow [1] as shown in the figure 1b.

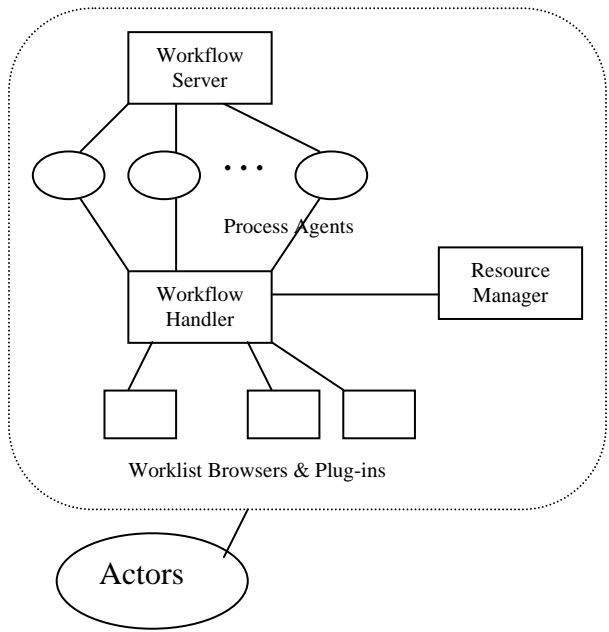

*Figure 1b: Architectural Diagrams of Workflow Management Systems [1]*

Suggested technique for the distribution of data in the distributed environment is *Distributing Only Static Data* i.e. the data which can persist only to that location rather then a variable or most changing data, to lessen the network traffic congestion. In that case only request has been made for the static data while variable data has to be operated locally [2].

We propose to use agent based technology to optimize the workflow by implementing a transactional like behavior among the tasks.

The next section will discuss the various synchronization and consistency issues among workflow tasks.

## III. Synchronization and Consistency related issues:

***Mismatched Data***: The data or information generated by one task must be of appropriate format for another task that requires it as an input, otherwise it will provide unacceptable results at later stages. Hence mismatched data should be identified and resolved.
***Inconsistent Data*:** In case of replicated data over a number of locations, the possibility of inconsistency exists. Hence adequate method has to be used to resolve the inconsistency.
***Completeness of Data***: Complete data should be available before the execution of the next task started. Hence a method is required to verify and ensure the completeness of the data.

## IV. Need of an existing research:

There is a need to avoid any flaw in the execution of the workflows by ensuring an efficient mechanism to bear the above mentioned issues and hence to make the execution flow successful. The workflow needs to be fault tolerant in case of hardware failure, but in this paper much of the consideration will be given to the execution of workflows on software grounds.

## IV. Proposed Solution:

Keeping in view the above-mentioned issues and objectives, an optimized solution is proposed that all the workflow activities within the workflow process will be bound with an agent that will ensure the correctness, consistency and completeness of data along with the synchronization among different tasks. The threads are also provided at different levels of an agent to do certain tasks simultaneously. This agent will be termed as synchronizing agent and is locally available to all workflow activities involved in a process. This means at each task level within the workflow process we have a synchronizing agent. Each synchronizing agent consists of three components as shown in figure 2a.

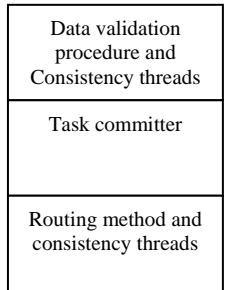

Figure 2a: Synchronizing Agent

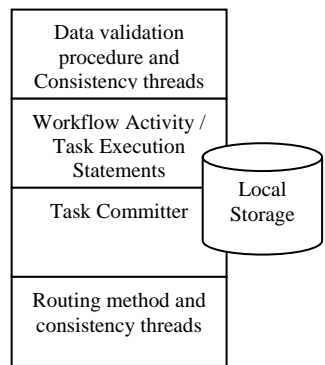

Figure 2b: Workflow Activity

Each workflow activity has locally available synchronizing agent and a local storage as shown in figure 2b.

**Components and working of synchronizing agent:**

*i) Data validation procedure and consistency threads:* It acts as the initialization condition for the workflow activity, Data validation procedure is used to check and ensure whether:
**a)-** The required data is completely available in the local storage associated with the task.
**b)-** The format of the required data is valid or not, if not valid then signals to the preceding task for valid format of data.
**c)-** When the task has to get required the data from two or more preceding tasks or sources then it waits to get the complete data.
**d)-** Once the complete data is available then the consistency check has been made to decide for the latest version of the data among multiple copies of same data and makes it available for use in the task. It also routes the selected copy of data to those tasks or sources that have inconsistent copy of that data. This routing is done with the help of consistency thread available.
**e)-** If the task requires the resources available locally then the above checks are bypassed.

Once the above conditions are satisfied then the execution of the task can be started.

*ii) Task committer:* It is a procedure used to check whether the current task has completed its execution successfully or not. There are two counter variables associated with the workflow task. One variable say "$t_e$" contained the total number of executable statements of that task and is available before the execution of that task. Another variable say "$t_{exec}$" is used to increase with the execution of each statement. Task Committer verifies the condition i.e. $t_{exec} < t_e$, if the condition yields true then it means that number of executed statements are less total number of executable statement. Hence the task could not be committed and the control has been shifted to the statement at offset given by $t_{exec}$. If condition yields false then this means the task is successfully executed and hence the task committer commits the task and signals the routing method that task has been completed.

*iii) Routing methods and consistency threads:* Once the routing method gets a signal from task committer then it signals the next coming task about its successful completion and sends all of the required data as requested by the next coming task at the time of loading that workflow process. The consistency threads at this level are responsible for making the data consistent by updating the inconsistent data with the acquired consistent data from the next task.

The workflow activity is initiated for the new task as described above.

**Components and working of Workflow Activities:**

The representation of workflow activity in figure 2b also follows ECA Rule [1, 9] i.e. Event- Condition-Action Rule as shown in figure 3a.

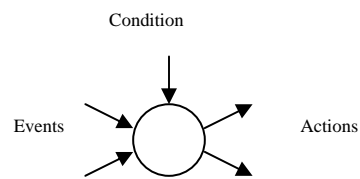

*Figure 3a: Workflow Activity under ECA Rule [9]*

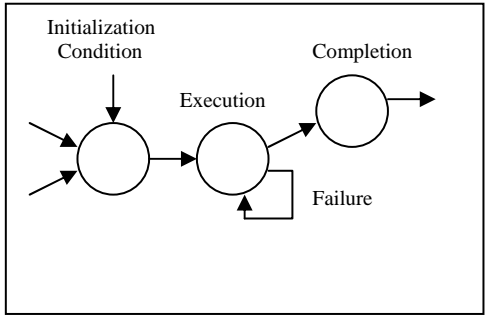

*Figure 3b: Internal States of workflow activity in the proposed solution*

Figure 3b shows the internal states in workflow activity that is started by initializing condition based on data validation procedure, then its execution started, if the execution is successful then task is committed other wise task committer will shift the control to the desired offset to continue. The committer counts the number of attempts made by the task to be committed. If the count exceeds a preset value, then the committer interrupts the workflow agent by sending a signal of execution failure. The agent sends that signal to the workflow server where possible hardware failure should be detected and alternative should be provided to resume the task execution. Finally the completion of the task is the last desirable state.

The process starvation or deadlock condition can be avoided by the use of Semaphores or Monitors within the tasks so that critical section can be locked to achieve mutual exclusion [7, 11].

## V. Model and Algorithm for workflow execution:

In this section, proposed algorithm is discussed, that includes three major steps required for execution of workflow in distributed system.

*Step 1: Loading and configuring a workflow process:*

The workflow process is configured in this phase, the configuration can be done again if the changes are made to the workflow process. As shown in figure 1, the process agents are responsible for taking the worklists to the workflow handlers. The following sub steps are involved to initialize the workflow engine for execution.

**i)** The number of instructions per task as mentioned by "$t_e$" has been calculated and mapped with that worklist. The workflow server also keeps a list of tasks and there number of executable statements in its local database.

**ii)** The synchronizing agent is bound with each of the workflow task as shown in figure 3b.

**iii)** NTP protocol [10] is applied to provide time synchronization among the tasks.

**iv)** The request for data to be used by a task is stored in the preceding task i.e. (Pre-fetch active)[2,6] for data requests. This will lower the network traffic at the time of execution of that workflow as one should not send requests for the data all the time it is required but the only emphasize will be to provide that data.

**iv)** Resource scheduling [7] for each task has been done, so that every task knows the sequence of resources it requires. For every resource the resource handler will need to contain the priority list of tasks that will consume a specific resource.

*Step 2: Execution of the workflow process:*

The following sub steps need to be followed to make sure the successful and optimized execution of the workflow process.

**i)** First task is ready for execution by having all the required data including its total number of instructions i.e. "$t_e$" available in its local storage.

**ii)** Data validation procedure activates and ensures the availability of correct, complete and consistent data, so that the task can start its execution.

**iii)** Once the data is validated then a counter variable as explained previously i.e. "$t_{exec}$" is initialized to zero.

**iv)** Now the task starts to execute and as the statements being executed then "$t_{exec}$" is increased at thread level.

**v)** Once the task completes its execution then the task committer will check whether the task is completely executed and commits that task.

**vi)** If the task is not completed then task committer directs the control to the place within that task from where the execution truncates. This would be done by getting the offset from "$t_{exec}$".

**vi)** If the task is unable to be completed in ten attempts then committer signals the workflow server to provide an alternate resource for the successful execution.

**vii)** Once the task is committed, then the activity would be in wait state for the control to be shifted to the next task. The routing method is called to fulfill the pre-fetch request for the data by sending the requested data to the next task. Once the acknowledgment [4] is received from the next task then the task gets into a completed state.

**viii)** The next task will perform the data validation step and if found multiple copies of same data then it will select the consistent data and the consistency threads at this level will transfer the consistent data to all those locations where the data needs to be updated to get consistency.

**ix)** If the format of the data is not appropriate the new task failed to start and signal the previous task to send the correct formatted data.

**x)** The task will only start its execution once it gets all of its requested data completely, correctly and consistently.

**xi)** The above mechanism will be repeated for all of the coming tasks and completed when there is no task left to execute. The complete execution is shown in figure 4.

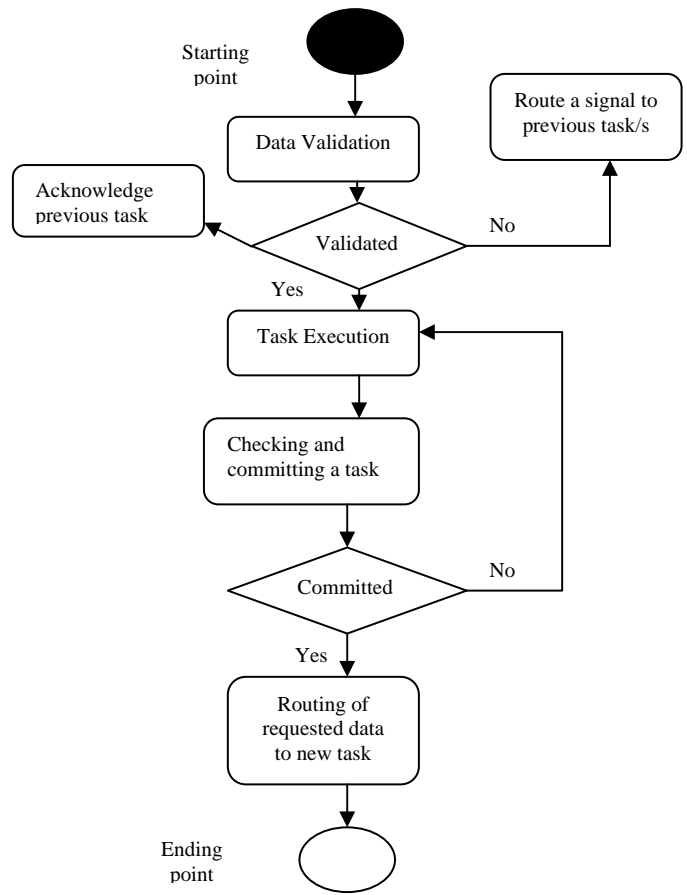

*Figure 4: Flow chart of the task execution*

### Step 3: Signaling the completion of the task to workflow server:

The successful completion of a workflow process is signaled to the workflow handler, which is then routed through the agent that initiates the process to the workflow server.

The above algorithm ensures the complete execution of a workflow process. A sample workflow process can be seen in the following figure 5 where A, B, C, D and E are Workflow activities.

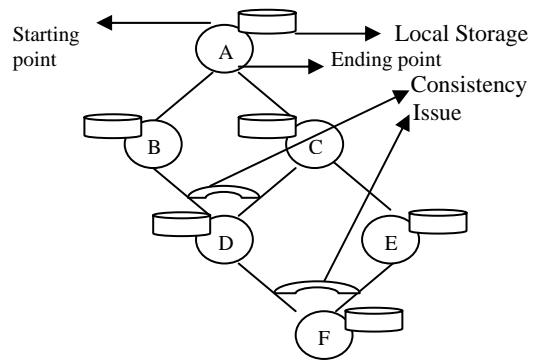

*Figure 5: Sample workflow*

## VI. Conclusion:

The proposed solution is very beneficial to acquire an optimized execution of a workflow process in distributed environment because process synchronization, data consistency and data correctness within the process is achieved at workflow task or workflow activity level. Moreover binding the synchronizing agent with the workflow task provides a fault tolerant and failure free execution of a workflow process as no task starts its execution with out the completion of preceding task and availability of required data. The model provides scalability to a workflow process as every new activity could be added to a process by updating only the list of tasks at workflow server. The whole process is synchronized due to the configuration done before the execution of the workflow process at both server and workflow task level. The solution behaves like a transaction as it does not allow any activity to start before committing the previous one but the thing that makes it more powerful is that, the whole activity does not need to repeat but the execution starts from the place where the task truncates in case if a task failed to commit. The model works well for the tasks having single or multiple instructions.

## VII. Limitations and Future Work:

There are few limitations and possible future work aspects in the proposed solution. The system is failure free but effective techniques to hardware failure must be incorporated to make the system robust, though one can detect the hardware failure using the count and acknowledgement mechanism provided at task level. As the task become heavy by the use of synchronizing agent even they are thread based, so there is a need to find out the ways to make these workflow tasks light weight. Another aspect to consider is that the proposed model works fine in the multi-instruction tasks but how can it be useful either in case of single-instruction task. More over the possible solution needs to be discovered regarding how the workflow process execution is made more efficient using the existing solution. It needs to be found out, how this solution is effective for the workflows in a grid environment.